\documentclass[10pt, conference]{IEEEtran}

\usepackage[T1]{fontenc}

\usepackage[utf8]{inputenc}

\usepackage{hyperxmp}

\usepackage{hyperref}
\hypersetup{
	pdfsubject={3rd IEEE Workshop on NEXt level of Test Automation 2020},
	pdftitle={Session-Based Recommender Systems for Action Selection in GUI Test Generation},
	pdfauthor={Varun Nayak, Daniel Kraus},
	pdfkeywords={Test generation, testing and debugging, information filtering},
	pdflang={en}
}

\usepackage{csquotes}

\usepackage[
	backend=bibtex,
	sorting=none,
	style=ieee,
	url=false,
	doi=false,
	isbn=false
]{biblatex}
\usepackage{siunitx}
\usepackage{graphicx}
\graphicspath{{resources/img/}}

\usepackage{booktabs}

\usepackage{paralist}

\usepackage{glossaries-extra}
\glsdisablehyper
\setabbreviationstyle[acronym]{long-short}

\newacronym{ai}{AI}{artificial intelligence}

\newacronym{ann}{ANN}{artificial neural network}

\newacronym{csv}{CSV}{comma-separated value}

\newacronym{etl}{ETL}{extract, transform, load}

\newacronym{gru}{GRU}{gated recurrent unit}

\newacronym{gui}{GUI}{graphical user interface}

\newacronym{ide}{IDE}{integrated development environment}

\newacronym{lstm}{LSTM}{long short-term memory}

\newacronym{ml}{ML}{machine learning}

\newacronym{mrr}{MRR}{mean reciprocal rank}

\newacronym{nlp}{NLP}{natural language processing}

\newacronym{rnn}{RNN}{recurrent neural network}

\newacronym{sut}{SUT}{system under test}

\usepackage[shortcuts]{extdash}

\renewcommand{\glsxtrfullsep}[1]{~}

\renewcommand*{\bibfont}{\footnotesize}

\newcommand{\squeezeup}[1]{\vspace{-#1mm}}

\begin{document}

\enquote{Session-Based Recommender Systems for Action Selection in GUI Test Generation} by Varun Nayak and Daniel Kraus, submitted to the 3rd IEEE Workshop on NEXt level of Test Automation (NEXTA) 2020. This is a preprint of the accepted version of this paper. The paper starts on the next page, after this information.

\copyright{}~2020 IEEE. Personal use of this material is permitted. Permission from IEEE must be obtained for all other uses, in any current or future media, including reprinting/republishing this material for advertising or promotional purposes, creating new collective works, for resale or redistribution to servers or lists, or reuse of any copyrighted component of this work in other works.

\clearpage

\title{Session-Based Recommender Systems for Action Selection in GUI Test Generation}

\author{
	\IEEEauthorblockN{Varun Nayak\IEEEauthorrefmark{1}, Daniel Kraus\IEEEauthorrefmark{2}}
	\IEEEauthorblockA{%
		ReTest GmbH \\
		Haid-und-Neu-Straße 7 \\
		76131 Karlsruhe, Germany \\
		Email: \IEEEauthorrefmark{1}\href{mailto:varun.nayak@retest.de}{varun.nayak@retest.de}, \IEEEauthorrefmark{2}\href{mailto:daniel.kraus@retest.de}{daniel.kraus@retest.de}}
}

\maketitle

\begin{abstract}
Test generation at the \gls{gui} level has proven to be an effective method to reveal faults. When doing so, a test generator has to repeatably decide what action to execute given the current state of the \gls{sut}. This problem of action selection usually involves random choice, which is often referred to as monkey testing. Some approaches leverage other techniques to improve the overall effectiveness, but only a few try to create human-like actions---or even entire action sequences. We have built a novel session-based recommender system that can guide test generation. This allows us to mimic past user behavior, reaching states that require complex interactions. We present preliminary results from an empirical study, where we use GitHub as the \gls{sut}. These results show that recommender systems appear to be well-suited for action selection, and that the approach can significantly contribute to the improvement of GUI-based test generation.
\end{abstract}


\glsresetall

\glsunset{csv}

\section{Introduction}

System tests through the \emph{\gls{gui}} are important since they stimulate software from end to end, i.e., somewhat from a user's perspective down to persistence layers such as databases. When used wisely, they can be a powerful part of a testing strategy. However, such tests usually have a bad reputation because they tend to be \textcquote{fowler12}{\textelp{} brittle, expensive to write, and time consuming to run.} Both academia and the industry try to overcome these issues by automatically generating GUI tests; not just to free developers and testers from the burden of test creation and maintenance, but to reduce the overall costs---without compromises and at the pace required~\cite{walgude19}.

While generated tests cannot fully compensate hand-crafted test cases, the future of testing is said to drastically increase the use of automation~\cite{wiklund18}. Nowadays, test generators already yield impressive results in a wide range of application areas. Sapienz~\cite{mao17}, for example, found \num{558} previously unknown crashes in an empirical study with more than \num{1000} Android apps from the Google Play store. Meanwhile, the former research project has been deployed at Facebook, where it is now used to automatically test the mobile apps of, e.g., Instagram, WhatsApp and Facebook itself~\cite{alshahwan18}.

When it comes to GUI-based test generation, a crucial part is to decide what action to execute next given the current state of the \gls{sut}. Many of today's approaches rely on random choice, a.k.a. \emph{monkey testing}. This is sometimes combined with techniques like (meta\=/)heuristics or \gls{ml} to improve the generated tests. For instance, ant colony optimization~\cite{bauersfeld11}, genetic programming~\cite{esparcia18}, \gls{ml}-enhanced evolutionary computing~\cite{kraus18}, data mining~\cite{ermuth16}, deep learning~\cite{li2019}, Q-learning~\cite{esparcia16} or other reinforcement learning algorithms~\cite{degott2019}. Yet, only a few actually focus on creating human-like sequences of actions, e.g., to allow a test generator to get behind \enquote{gate \glspl{gui}}~\cite{amalfitano19} such as login screens or non-trivial forms. And although random testing is effective in finding relevant faults, it tends to miss bugs humans do reveal~\cite{ciupa08}. Therefore, generating sequences that mimic past user behavior might help to reduce this gap.

We propose a novel approach to action selection in GUI-based test generation by leveraging \emph{recommender systems}. Recommender systems are a well-studied field and they form the core of many successful businesses like Netflix~\cite{gomez16} or YouTube~\cite{covington16}, for which targeted recommendations are indispensable. We investigate a possible intersection between recommender systems and the problem of action selection by mapping \gls{gui} actions to items and sessions within a SUT to users. Provided an adequate amount of data, our approach is able to predict actions a user likely would perform in the current state. By using a \emph{session-based} recommender system as our model, we not just suggest single actions, but sequences of actions. This allows a test generator to be guided through states that require complex user interactions.

First, we give a brief introduction to recommender systems and some advances relevant for this paper in Section~\ref{sec:basics}. Section~\ref{sec:approach} outlines our technical approach, where we describe the overall design and various implementation details. In Section~\ref{sec:eval}, we conduct a first empirical study on top of GitHub by mixing real-world and synthetic data. Afterwards, we summarize our findings and report on our planned future work in Section~\ref{sec:summary}.

\section{Recommender Systems}
\label{sec:basics}

Receiving recommendations of different forms has become a part of our daily online experience in a variety of application domains such as e-commerce, social media and content streaming. Internally, such systems analyze the past behavior of individual users to detect patterns in data. On typical online sites, various types of user actions can be recorded, e.g., that a user views an item or makes a purchase. These recorded actions and the detected patterns are then used to provide recommendations to the user. In this context, the entity being recommended is called \emph{item}, and the entity that receives the recommendation is referred to as the \emph{user}.

The basic models for recommender systems work primarily with user-item interactions such as ratings or like/dislike, or attributes like user interests or item properties. Based on this, there are two main approaches to recommender systems: \emph{collaborative filtering} and \emph{content-based}. Traditional techniques such as matrix factorization have treated user-item interactions as flat, matrix-structured data, often ignoring the temporal structure and order within the data~\cite{wang19}. Being able to predict a user's short-term interests in an online session is a highly relevant problem in practice, e.g., to adapt to item viewing and purchase activities in e-commerce. Within such application domains, the items have to be recommended in a certain order, or the recommendation of one item only makes sense after some other event has happened.

\emph{Session-based} recommender systems consider the information embedded in between sessions and treat sessions as the basic recommendation unit. A session could be a set of items or a collection of actions consumed in one event or in a particular period of time. When dealing with such sequential data, \glspl{rnn} are being heavily used~\cite{lipton15}. But practical applications involve temporal dependencies spanning many time steps where the network is often unable to propagate useful information from the output end of the model back to the layers near the input end, known as the vanishing gradient problem~\cite{bengio94}.

In 2016, Hidasi et al.~\cite{hidasi16b} presented a \emph{\gls{gru}}-powered \gls{rnn} for session-based recommendations and called this method \emph{GRU4Rec}. A \gls{gru} is a more elaborate model of an \gls{rnn} that can deal with the aforementioned problem(s)~\cite{cho14}. The gatings within these units essentially learn when and how much to update the hidden state of the unit. This enables more accurate recommendations for session-based data.

\section{Our Technical Approach}
\label{sec:approach}

\begin{figure}
	\centering
	\includegraphics[width=0.5\columnwidth]{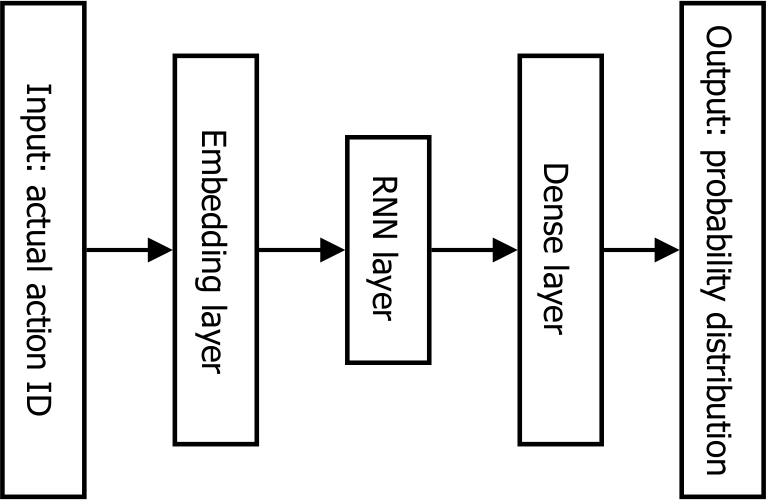}
	\caption{General architecture of our network based on \cite{hidasi16b}.}
	\label{fig:architecture}
	\squeezeup{3.5}
\end{figure}

We design our approach on top of the work by Hidasi et al. as illustrated in Figure~\ref{fig:architecture}. We adopt their general network architecture, but specialize it for our purposes.

The input to our model is a batch of sessions where each session is encoded as a sequence of action IDs. Action IDs are derived from the targeted \gls{gui} element and the action performed on it. That is, two actions only produce the same ID if they target the same element with the same action (ignoring possible input data like text). An \gls{rnn} layer is added next, which consists of \gls{gru} or \gls{lstm}. (In our evaluation below, we explore multiple network types.) This layer is expected to learn the temporary patters in the action-selection behavior. The following dense layer converts this information into a probability distribution over the given action IDs---the output of our model. Thus, we get a stream of ranked actions that is based on past behavior of actual users a test generator can choose from.

To implement our approach, we are using the GRU4Rec library based on \cite{hidasi16b, hidasi17}. It adds several extensions to a sequence modeling architecture like ours. For example, session-parallel mini-batches, mini-batch-based output sampling and the use of a pairwise ranking loss function. Sequence-to-sequence models produce the result by one item at a time, in other words, by solving a classification problem at each time step. According to Hidasi et al., pairwise ranking losses are expected to give a better performance with the given network setup. The loss function compares the rank of pairs of a positive and a negative item and enforces that the rank of the positive item should be lower than that of the negative one.

For more details on the basic network setup, please refer to the original paper(s). Next, we conduct a first empirical study, show how the network can be trained and how it performs.

\section{A First Empirical Evaluation}
\label{sec:eval}

According to the World Quality Report 2018-19~\cite{buenen18}, data is a main obstacle when it comes to the adoption of \gls{ai} in testing. The problem of data scarcity is an important factor since data is at the core of any \gls{ml} project. This issue is especially challenging for young or small organizations, because only rarely there are cooperations with large enterprises where sufficient data is available.

As we were struggling with small data too, we have designed an \gls{etl} pipeline to increase the amount of real-world data by adding synthetic data, so that we get a first impression of how our approach performs. We decided to target web applications (i.e., web-based \glspl{gui}), where we picked GitHub as the \gls{sut}; a popular code hosting and development platform. This setup allows insights based on
\begin{inparaenum}[(i)]
	\item a mature and widely-used target platform,
	\item a sufficiently complex and well-known \gls{sut}.
\end{inparaenum}

To create real-world data, we recorded \num{50} of our own user sessions on GitHub using the Selenium IDE, a record-and-playback tool available as a Chrome and Firefox extension. We exported these sessions as Java tests, where every test case represents a user session. Each exported test was executed with a custom Selenium WebDriver, which allows us to extract training data as \gls{csv}. Note that the Selenium IDE comes with the functionality that when the default locator---a particular \gls{gui} element property such as an ID, typically used in test scripts to locate elements---doesn't find an element, it will fall back to other available means. This fallback mechanism ensures that most recorded tests don't fail, e.g., due to recording inaccuracies. The exported Java code does not have this feature and only uses a single locator, which is why we had to manually fix many locators to avoid runtime failures. The poor code export quality when using the Selenium IDE is a major bottleneck in the proposed \gls{etl} pipeline that we aim to address as part of our future work (see Section~\ref{sec:summary}).

\begin{table}
	\centering
	\caption{Real-world data samples after pre-processing.}
	\label{tab:data-samples}
	\begin{tabular}{lll}
	\toprule
	Session ID & Action IDs sequence               & Timestamp  \\
	\midrule
	1          & (151, 1, 2, 3, 4)                 & 1568573073 \\
	2          & (151, 4, 5, 3, 1, 2, 3, 4)        & 1568573079 \\
	3          & (6, 7, 8, 9, 10, 11, 12, 2, 3, 4) & 1568573088 \\
	4          & (151, 4, 5, 3, 4, 1, 2, 3, 4)     & 1568573099 \\
	5          & (6, 109, 110, 2, 3, 4)            & 1568573362 \\
	\bottomrule
	\end{tabular}
	\squeezeup{3.5}
\end{table}

One of the key parts in the pre-processing step is the assignment of accurate action IDs. The tests exported via the Selenium IDE already contain the absolute XPath for each element, we combine this information with the web page the element appears on and the performed action type to derive the action ID. Table~\ref{tab:data-samples} illustrates some resulting data samples, where every session is of arbitrary length (the actions executed by the user) and represented by a sequence of action IDs.

\begin{figure}[h]
	\centering
	\includegraphics[width=\columnwidth]{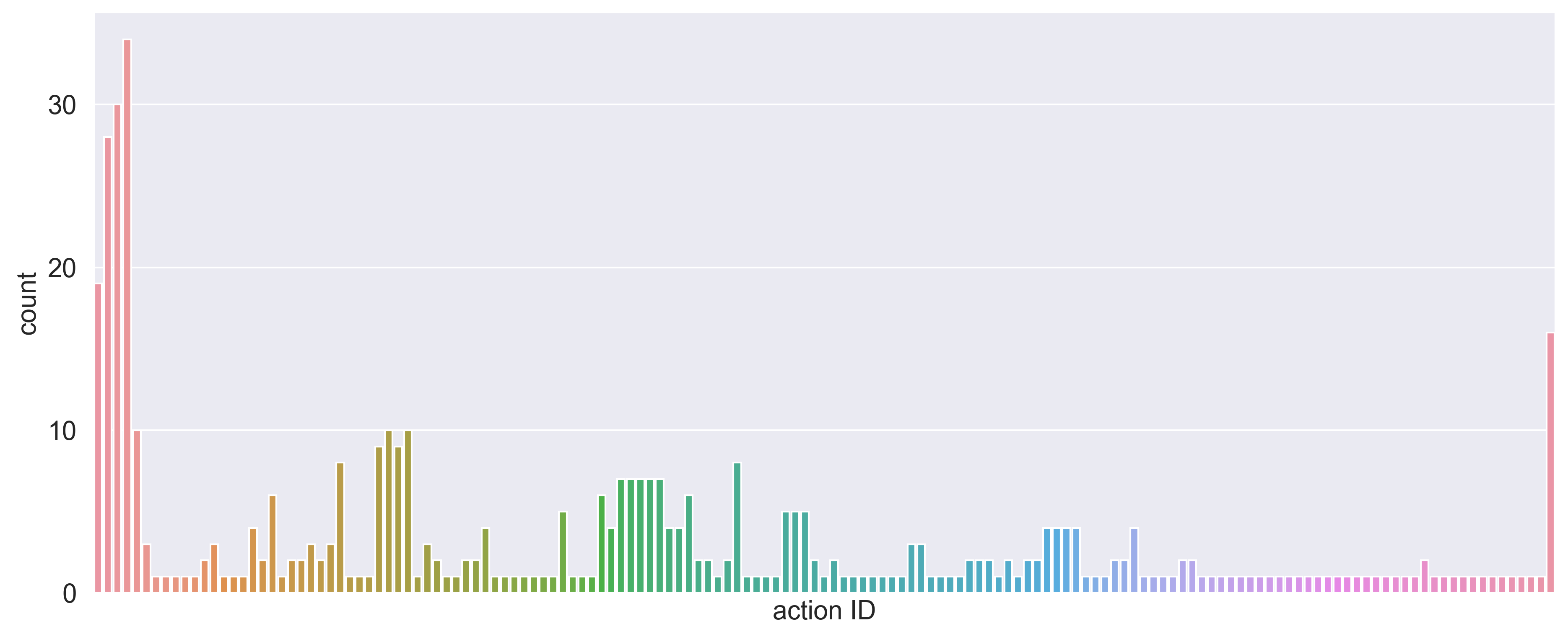}
	\caption{Action ID distribution within the recorded user sessions.}
	\label{fig:action-id-dist}
\end{figure}

When it comes to synthetic data, Wu et al. \cite{wu16} formalize the problem of generating such datasets using the maximum entropy principle for categorical data, which captures the characteristics of the underlying data. Figure~\ref{fig:action-id-dist} shows the distribution of action IDs within the recorded user sessions. As can be seen, some actions are more common across sessions. These frequent actions usually carry a deeper meaning and represent short-term user goals that correspond to common use cases like a login procedure. The assumption we made is that most sessions will have such recurring patterns in the recorded interactions and in-between a user will perform arbitrary actions. The synthetic sessions we generated still hold these properties, but have been mixed up with randomly created action IDs. In practice, click botnets may also create a considerable amount of traffic~\cite{nagaraja19}, so we additionally interjected random noise to represent spam and the like.

The resulting dataset is summarized in Table~\ref{tab:dataset-summary}. We split this data into a training set (\SI{80}{\percent}) and a test set (\SI{20}{\percent}) for evaluation. In the context of recommender systems, we are most likely interested in recommending an item from the top-$n$ list of items. Therefore, we calculate relevant metrics with regards to the first $n$ actions instead of all actions. Precision at $n$ is the proportion of recommended actions present in the top-$n$ list that are relevant. Recall at $n$ is the proportion of relevant actions found in the top-$n$ recommendations. \Gls{mrr}, which is important in cases where the order of recommendations matter, is the mean of reciprocals of the rank from all queries. The reciprocal rank is set to zero if the rank is above $n$.

\begin{table}
	\centering
	\caption{Summary of the used dataset.}
	\label{tab:dataset-summary}
	\begin{tabular}{lrllr}
	Real-world sessions & \num{50}   & & Avg. no. of actions & \num{14.12} \\
	Synthetic sessions  & \num{3476} & & Min. no. of actions & \num{1} \\
	Distinct actions    & \num{522}  & & Max. no. of actions & \num{49}
	\end{tabular}
	\squeezeup{3.5}
\end{table}

We further followed the practice of Quadrana et al.~\cite{quadrana18}, where items from each session in the test set are grouped together to form a sequence, and each sequence is further split into the user profile and ground truth. The user profile is composed of the first event in the sequence that is fed into the system and used to compute recommendations. The ground truth is composed of the remainder of the sequence that is used for performance evaluation. Items are revealed incrementally, then the evaluation is performed after each new item. This helps to evaluate the recommendation quality in a setting where user profiles are revealed sequentially. Metrics are averaged over each sequence and then averaged over all.

\begin{figure}[h]
	\centering
	\includegraphics[width=\columnwidth]{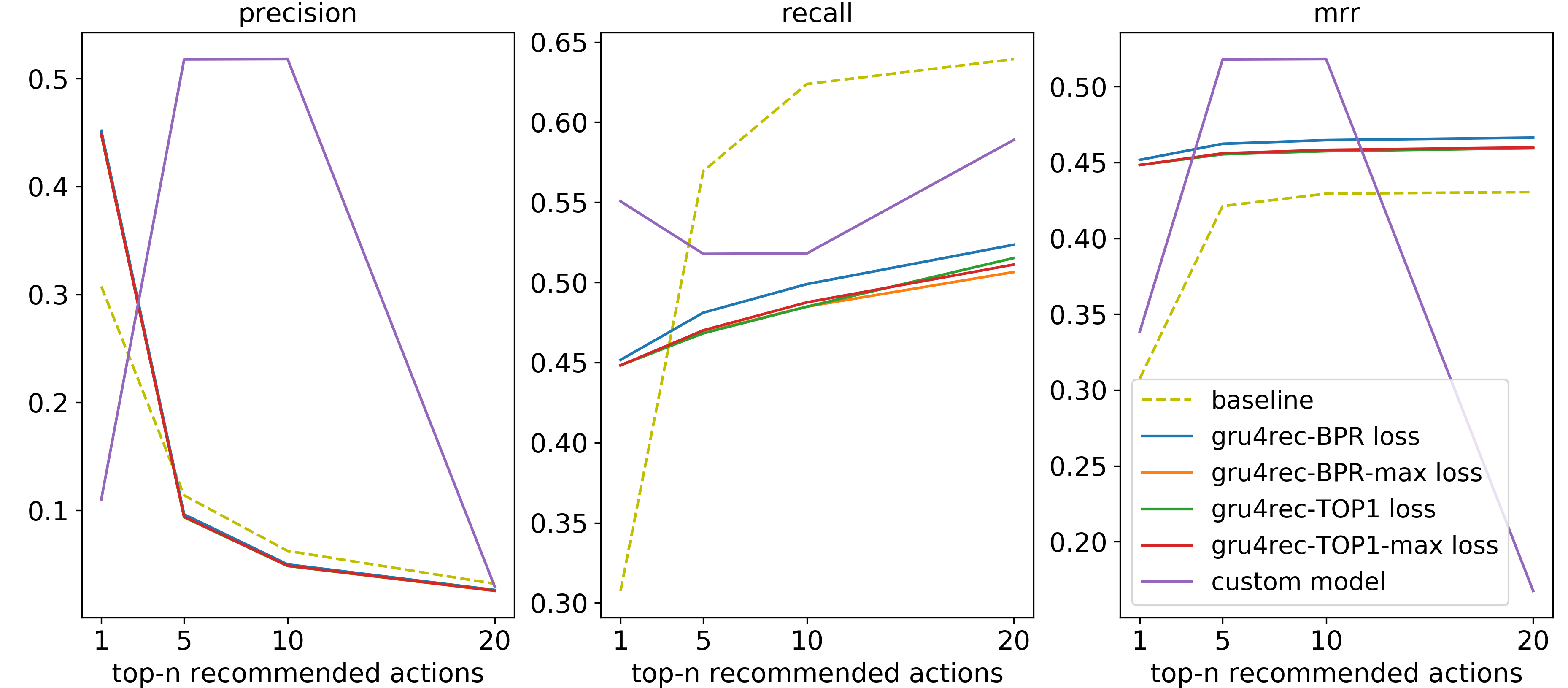}
	\caption{Precision, recall, MRR at 1, 5, 10, 20 for the different network types.}
	\label{fig:results}
\end{figure}

As a baseline, we used a simple $k$-nearest neighbor recommender based on an item-to-item similarity. In this setting, the similarity matrix is pre-computed from the available session data, i.e., actions that are often executed together in sessions are deemed to be similar. This similarity matrix is then used during the session to recommend the most similar actions to the one the user has currently performed. We compared this model to different \gls{gru}-based network types with different losses, as well as a custom model using \gls{lstm} and cross-entropy loss. The models converged between \numrange{5}{25} epochs, depending on the loss function and the amount of data. The evaluation results are shown in Figure~\ref{fig:results}.

An observation we made is that the \gls{mrr} is roughly within the range $[0.2, 0.5]$ across all the evaluated recommendation list lengths. This indicates that the best relevant actions were retrieved between top-5 and top-10. The \gls{mrr}, however, seems to saturate, which we assume is a consequence of the data deficit. The baseline performance being on-par with the other models is most likely owing to short average session length. Precision and recall both indicate the accuracy of the models. Precision for all models except the custom one is very high at the top-1 recommendation, then it continues to drop as the recall climbs up. This is because in order to recall everything, it is required to keep generating results which are not accurate, hence, lowering the precision.

We also observed that the models did not significantly outperform the baseline in the top-5 recommendation and beyond. In all experiments, the best-choice model performed better only by \SIrange{8}{18}{\percent}. This could possibly have to do with the used dataset, but the lack of data remains a threat to validity and adds some uncertainty. Apart from this, the models performed well. The recommended sequences reflect many of the recorded use cases, mastering also complex states.

\section{Conclusion and Future Work}
\label{sec:summary}

We have built a novel prototype for action selection in \gls{gui} test generation using a session-based recommender system. We conducted a first empirical study on top of GitHub, for which we presented preliminary results. These results suggest that action selection, when seen as a sequence modeling task, can guide a test generator through states that require complex interactions by mimicking past user behavior.

Based on our current approach and findings, we identified several tasks for future work. First, there is an overall need for more real-world data in order to develop sophisticated models. Therefore, we strive for cooperations with owners of big web applications and other researchers. Second, a major bottleneck was the poor quality of the Selenium IDE code export. We plan to develop
\begin{inparaenum}[(i)]
	\item a native Selenium IDE plugin to leverage fallback locators and
	\item a script for web application owners to extract anonymous usage traces.
\end{inparaenum}
Third, adopting additional GRU4Rec extensions from \cite{hidasi17} could improve the results. Moreover, hyperparameter optimization tools may be used to further improve the models' performance. Last and most importantly, the presented results are preliminary. To prove its effectiveness, the approach must be evaluated as part of a large-scale study using multiple, diverse \glspl{sut}, ideally in comparison to other test generators.

We believe that addressing these tasks can significantly contribute to the improvement of \gls{gui}-based test generation.

\section*{Acknowledgements}

As part of the joint research project \enquote{Surili}, this work is supported by a grant (no. 01IS17092A) from the German Federal Ministry of Education and Research.

\printbibliography

\end{document}